# Interface Effects on the Ionic Conductivity of DopedCeria – Yttria-stabilized Zirconia Heterostructures

*Daniele Pergolesi[1*], Elisa Gilardi[1], Emiliana Fabbri[1], Vladimir Roddatis[2], George F. Harrington[3,4,5], Thomas Lippert[1,6,7], John A. Kilner[3,7] and Enrico Traversa[8,9]*

[1] Research with Neutron and Muon Division, Paul Scherrer Institut, 5232 Villigen-PSI, Switzerland

[2] Institut für Materialphysik, Universität Göttingen, 37077 Göttingen, Germany

[3] Department of Materials, Imperial College London, London SW7 2BP, United Kingdom

[4] Next-Generation Fuel Cell Research Centre, Kyushu University, 744 Motooka, Nishi-ku Fukuoka 819-0395, Japan

[5] Department of Materials Science and Engineering, Massachusetts Institute of Technology, 77 Massachusetts Ave., Cambridge MA 02139, U.S.A.

[6] Department of Chemistry and Applied Biosciences, Laboratory of Inorganic Chemistry, Vladimir-Prelog-Weg 1-5/10, ETH Zürich, 8093 Zürich, Switzerland

[7] International Institute for Carbon Neutral Energy Research (WPI-I2CNER), Kyushu University, 744 Motooka, Nishi-ku, Fukuoka 819-0395, Japan

[8] International Research Center for Renewable Energy, Xi'an Jiaotong University, Xi'an, Shaanxi 710049, P. R. China

[9] NAST Center & Department of Chemical Science and Technology, University of Rome Tor Vergata, 00133 Rome, Italy

\* daniele.pergolesi@psi.ch

**Abstract**

Multilayered heterostructures of $Ce_{0.85}Sm_{0.15}O_{2-\delta}$ and $Y_{0.16}Zr_{0.92}O_{2-\delta}$ of a high crystallographic quality were fabricated on (001) - oriented MgO single crystal substrates.
Keeping the total thickness of the heterostructures constant, the number of ceria-zirconia bilayers was increased while reducing the thickness of each layer. At each interface Ce was found primarily in the reduced, 3+ oxidation state in a layer extending about 2 nm from the interface. Concurrently, the conductivity decreased as the thickness of the layers was reduced suggesting a progressive confinement of the charge transport along the YSZ layers. The comparative analysis of the in-plane electrical characterization suggests that the contribution to the total electrical conductivity of these interfacial regions is negligible. For the smallest layer thickness of 2 nm the doped ceria layers are electrically insulating and the ionic transport only occurs through the zirconia layers. This is explained in terms of a reduced mobility of the oxygen vacancies in the highly reduced ceria.

**Introduction**

Yttria-stabilized zirconia (YSZ) with a typical composition of 8 mol.% $Y_2O_3$ doped $ZrO_2$ and Gd or Sm-doped ceria (GDC and SDC) with typical dopant content of 7.5 or 10 mol.% are oxygen ion conducting oxides that find their most important application as electrolyte materials for solid oxide fuel cells (SOFCs). Currently commercialized technology is based on an anode-supported design using YSZ as the electrolyte and a typical operating temperature above 900°C which is required to achieve acceptable ohmic losses across the YSZ membrane. Such a high operating temperature is a severe drawback for the chemical and mechanical stability of the cell and enormous advantages would arise from the possibility of preserving good cell performance in the so-called intermediate temperature range, between 600 and 800°C.[1]

Another important issue when YSZ is used is its chemical reactivity with the best performing cathode materials. These are mixed ionic and electronic conductors, such as $La_{1-x}Sr_xCo_{1-y}Fe_yO_{3-\delta}$ (LSCF),[2,3] which extends the electrochemically active area (where the oxygen reduction reaction takes place) to the entire cathode surface. The reactivity of LSCF with YSZ starts around 1000°C and results in the formation of ionically insulating Sr and La zirconates at the electrolyte-cathode interface.[4-7]

Compared to YSZ, doped ceria offers significantly higher conductivities at lower temperatures due to the lower activation energy for oxygen ion migration (around 0.7 eV compared to about 1 eV for YSZ). Moreover, doped ceria would make it possible to use the best cathode materials currently available since coupling with LSCF is chemically stable.[4,6,8] Ceria-based materials are also beneficial at the anode side decreasing polarization losses.[9] However, at low oxygen partial pressure (specifically at the anode side of the cell) the reduction of $Ce^{4+}$ to $Ce^{3+}$ introduces a significant electronic contribution to the conductivity precluding the use of this material as a relatively thin electrolyte membrane.

Strategies have been proposed such as using doped ceria at the anode side, coupled with erbia-stabilized bismuth oxide at the cathode,[10] or doped ceria at the cathode side of a YSZ electrolyte as a diffusion barrier interlayer to avoid chemical reaction between YSZ and LSCF.[4,6,8,11] The effectiveness of the latter depends on the coating technology. When wet ceramic processes are used, a sintering step at temperatures as high as 1300 °C is required to obtain fully dense layers with good Sr and La ion retention. In this temperature range however, ceria and zirconia diffuse into each other forming solid solutions with low conductivity. Thin film deposition technologies, such as sputtering, allow the growth of fully dense doped ceria layers at much lower temperatures (700-800°C).[4, 11] The subsequent co-sintering process required for the preparation of the LSCF cathode is performed around 1100°C, a temperature range where the formation of a ceria-zirconia solid solution is still not favoured, and the reported fuel cell tests are very promising.[4, 6, 8]

YSZ and doped ceria thin films have also been used to probe the effect of stress-induced lattice distortions (strain) on the ionic conductivity, a topic of intense investigation for almost a decade.[12] [13] The interest is justified by several experimental as well as theoretical studies that show and predict increased conductivity in tensile strained lattices, while the opposite effect is expected as a result of compressive strain.[14-21] To investigate these effects, thin films of ionic conducting materials have been fabricated both as single layers on single crystal substrates or as multilayered heterostructures coupling the oxygen ion conductor under investigation with a more insulating material with a different lattice parameter. In both cases, the strategy is to produce an elastic strain arising from the stress induced at the interface by the different lattice parameters of the two materials, thus allowing in-plane conductivity measurements (along the direction of the substrate surface) of the ionic conductors under bi-axial strain.

In a different approach, thin film multilayered heterostructures fabricated by coupling two oxygen ion conductors, typically using SDC or GDC and YSZ or gadolinia-stabilized zirconia (GSZ), have been used to investigate the effect of the respective lattice distortions on the total conductivity.[22-23] Following this strategy, the interpretation of the experimental results is in general more complicated. Due to the lattice mismatch of the two materials, the doped ceria layers are expected to develop an in-plane compressive stress, while an in-plane tensile stress is expected for the zirconia layers. It is thus difficult to predict the total effect resulting from the two opposite trends. These systems however are a powerful tool to achieve more insights into the physicochemical properties of the interface between the ceria and zirconia based materials, a scientific target of important technological interest.

Here we investigate the structural, chemical and conducting properties of the SDC-YSZ interface using epitaxial highly ordered thin film multilayered heterostructures. At constant total thickness, the number of symmetric SDC-YSZ bilayers is increased while reducing the thickness of the layers. The good crystalline quality of the films allowed the fabrication of a reference sample as a fully relaxed bilayer of the two materials showing the expected conductivity of two YSZ and SDC single crystals in parallel configuration. The conducting properties of the interface become progressively evident while reducing the thickness of the layers.

**Experimental**

Thin film heterostructures of 7.5 mol % $Sm_2O_3$ doped $CeO_2$ (SDC) and 8 mol % $Y_2O_3$ stabilized $ZrO_2$ (YSZ) were fabricated by pulsed laser deposition (PLD). Commercially available (001)-oriented MgO single crystal wafers were used as substrates. Epitaxially oriented heterostructures were obtained using a double buffer layer of $BaZrO_3$ (BZO) and $SrTiO_3$ (STO), as described in [24].

All layers were grown under the same deposition parameters as follows: target to substrate distance of 60 mm, oxygen partial pressure of about 3 Pa, laser energy at the target of about 32 mJ on a spot area of about 2 mm$^2$ (energy density of about 1.6 J/cm$^2$), heater set temperature of 800°C. Platinum paste was used to provide the thermal contact between the substrate and the heater. The substrate temperature was estimated to be about 750°C. The samples were post-annealed *in situ* for 15 minutes at an oxygen partial pressure of 75 Pa before cooling down to room temperature at 5°/min.

The targets of SDC and YSZ were fabricated by an assisted coprecipitation method from nitrate water solutions as reported in [25]. The powders were uniaxially pressed at 140 MPa and sintered at 1450 °C for 10 hours.

X-Ray Diffraction and X-ray Reflectometry (PANalytical X'pert Pro MPD with Cu K$\alpha_1$ radiation at 1.540Å) analyses were used to investigate the crystalline structure of the films and for the calibration of the deposition rate, respectively. With the selected deposition parameters and geometry, a very similar deposition rate was found for BZO, STO and SDC, while approximately a factor of 3 lower deposition rate was found for YSZ. Each material was deposited at a rate of about 0.2 Å/s using a laser frequency of 1 Hz for BZO, STO and SDC, while 3 Hz was used for YSZ.

The growth of the heterostructures was monitored *in situ* by high-pressure reflection high-energy electron diffraction (RHEED).

High-resolution scanning transmission electron microscopy (HR-STEM) was carried out using a FEI Titan 80-300 environmental TEM equipped with a high-angle annular dark-field (HAADF) detector, operated at 300 kV. Electron energy loss spectroscopy (EELS ) was performed using a Gatan GIF Quantum 965 ER. The samples were prepared by standard techniques, including mechanical polishing followed by ion milling using a PIPS 695 ion mill.

Microstructural analysis and atomic energy dispersive spectroscopy (EDX) were performed to check the distribution of the different elements along the layers. This was carried out on a JEOL JEM-ARM200F transmission electron microscope, with a SDD type X-ray detector, operating at 200 kV. A cross sectional sample was prepared using a focused ion beam (FIB) system (FEI

Helios NanoLab). The sample was prepared and thinned at 30 kV and then subsequently at 5 kV and 2 kV to minimize the amorphous damage layer.

For electrical characterization, two Pt films about 100 nm thick were deposited as electrodes on the film surface by electron beam deposition at room temperature. 5 nm thick Ti layers were used to improve the adhesion between the Pt and the sample. The distance between the electrodes was about 1 mm. Impedance spectroscopy (IS) measurements were performed using a multichannel potentiostat VMP3 (Bio-Logic) between 100 mHz and 1 MHz in air, varying the temperature between 325 and 680°C. Electrical measurements at low oxygen partial pressure were performed at 700 °C using a gas mixture of 5% $H_2$ in $N_2$. The oxygen partial pressure was estimated from a previous calibration with an oxygen sensor.

**Results and Discussion**

Five sets of epitaxial heterostructures of 7.5 mol % Sm-doped $CeO_2$ (SDC) and 8 mol.% Yttria-stabilized zirconia (YSZ) were grown on (001)-oriented MgO single crystal substrates, varying the layer thicknesses and the number of interfaces.

**Table 1**. Sample design: total thickness 200 nm, BZO and STO thicknesses 7 nm.

| MgO + BZO + STO + (SDC+YSZ) × n | |
|---|---|
| Thickness SDC+YSZ (nm) | Number bilayers N |
| (50 + 50) | 2 |
| (20 + 20) | 5 |
| (10 + 10) | 10 |
| (5 + 5) | 20 |
| (2 + 2) | 50 |

The described substrate design (MgO + BZO + STO), shows very low total conductivity making this template platform suitable for in-plane electrical characterization of thin SDC and YSZ films, as discussed in detail elsewhere.[24]

The thickness of the two layers of BZO and STO grown in this sequence was about 7 nm. The five sets of heterostructures were fabricated with the same total thickness of about 200 nm but with a different number of SDC + YSZ bilayers of different thickness (with the same thickness for the two materials in each bilayer), as reported in Table 1.

**Structural, morphological, and chemical characterization**

Figure 1 shows the results of the X-ray diffraction (XRD) analysis of the heterostructures. In the 2θ−ω scans of Figure 1a, the (002) reflex of BZO is visible as a shoulder at the base of the substrate peak at slightly larger 2θ angles. The (002) reflex of STO is also visible in the expected angular region, though the small thickness of the film (5 - 7 nm) broadens the diffraction peak. Interference fringes are visible within the angular region comprising the (002) reflexes of MgO, BZO, and STO indicating the good quality of the interfaces.[19]

In the angular region between 2θ values of 32° and 36° the XRD features of the (002) reflexes of YSZ and SDC can be observed. This out-of-plane XRD analysis is very similar to that reported for similar samples in reference.[19]

For the (50+50)×2 sample the (002) diffraction peaks of the two materials are well resolved and from their angular positions the almost fully relaxed out-of-plane lattice parameters of about 5.43 and 5.14 Å can be calculated for SDC and YSZ, respectively. When the thickness of each layer is reduced the interference effect due to the increasing number of interfaces hinders the identification of the angular position of the diffraction peaks. For layer thicknesses below 10 nm the XRD plots in Figures 1a and 1b show the typical superlattice features where an average structure peak ($SL_0$), as well as $1^{st}$ and $2^{nd}$ order satellite peaks ($SL_{\pm n}$) can be identified. The presence of satellite peaks near the main Bragg peak reflects the good crystalline quality of the heterostructures with well-defined interfaces. The angular positions of the satellite peaks allow the calculation of the thickness, $d$, of the SDC + YSZ bilayer in the superlattice given by $d = n\lambda/(\sin\vartheta_{-n} - \sin\vartheta_n)$, where $\lambda$ is the X-ray wavelength and $\vartheta_{\pm n}$ is the angular position of the $n^{th}$ order satellite peaks. A bilayer thickness of 9.63 and 4.81 nm can be estimated for the samples (5+5)×20 and (2+2)×50, respectively. These values are in good agreement with the expected deposition rate calibrated by X-ray reflectometry.

Reciprocal space maps (RSM) measured for the (5+5)×20 and (2+2)×50 superlattices are shown in Figures 1c and d, respectively. The out-of-plane reciprocal vector $Q_z$ of the sample (5+5)x20 shows the typical superlattice features equivalent to those observed in the 2θ-ω scan, while for the sample (2+2)x50 the intensity of the satellite peaks was too small to be detected by RSMs. In the RSM plots the red squares indicate the values $Q_x$ and $Q_z$ of the reciprocal vectors of the two materials in the case of the fully relaxed crystalline structures. For both heterostructures the region of maximum intensity of the in-plane component of the reciprocal vector $Q_x$ stretches from about 1.65 to 1.72 Å$^{-1}$. The in-plane lattice parameters of fully relaxed SDC (a=5.43 Å) and YSZ (a=5.14 Å) are outside such a range, the value being larger for SDC and smaller for YSZ. This suggests the presence of an in-plane compressive strain for SDC and tensile for YSZ, though due to the large lattice mismatch we expect the stress to be largely released by the introduction of misfit dislocations. Any attempt to quantify the residual and effective strain from these measurements would be merely speculative.

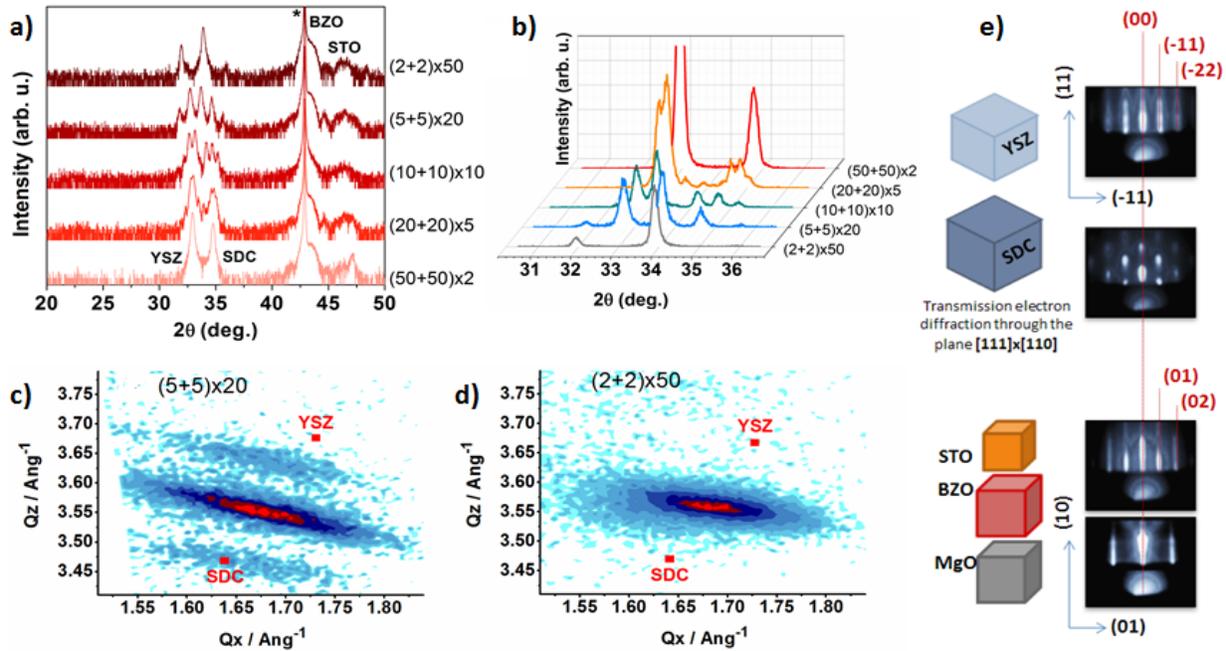

**Figure 1.** a) Representative ω/2θ scans of the heterostructures identified by (m+m)×n, where m is the thickness of the layers and n is the number of SDC + YSZ bilayers. The substrate peak is marked with *. b) ω/2θ scan in the angular region comprising the (002) reflexes of YSZ and SDC (intensity is on a linear scale). c) and d) Reciprocal space maps of the (5+5)×20 and (2+2)×50 heterostructures acquired along the (113) asymmetric diffraction peaks of SDC and YSZ. The red squares indicate the reciprocal vectors of the fully relaxed crystal structures of SDC and YSZ. e) Representative RHEED analyses acquired during the growth of the different layers.

The growth of the heterostructures was monitored *in situ* by RHEED. The epitaxial relations of the various layers is schematically shown in Fig. 1e together with representative RHEED patterns that confirmed the growth mechanism already reported in the literature.[19] A 2-dimensional layer-by-layer growth of the two buffer layers is followed by a 3-dimensional growth (island-like surface morphology) of the SDC layer. A 2-dimensional surface reconstruction is observed during the growth of the YSZ layer that reduces the surface roughness. The following SDC/YSZ layers reproduce the same sequence of 3-dimensional and 2-dimensional growth for all the heterostructures. We observed that the 2-dimensional surface reconstruction of the YSZ layer is evidently recognizable in the RHEED pattern for YSZ layer thicknesses larger than about 2 nm. For what follows it is important to highlight that a coherent diffraction pattern was always present during the growth of the samples and in particular at the interface between two different materials.

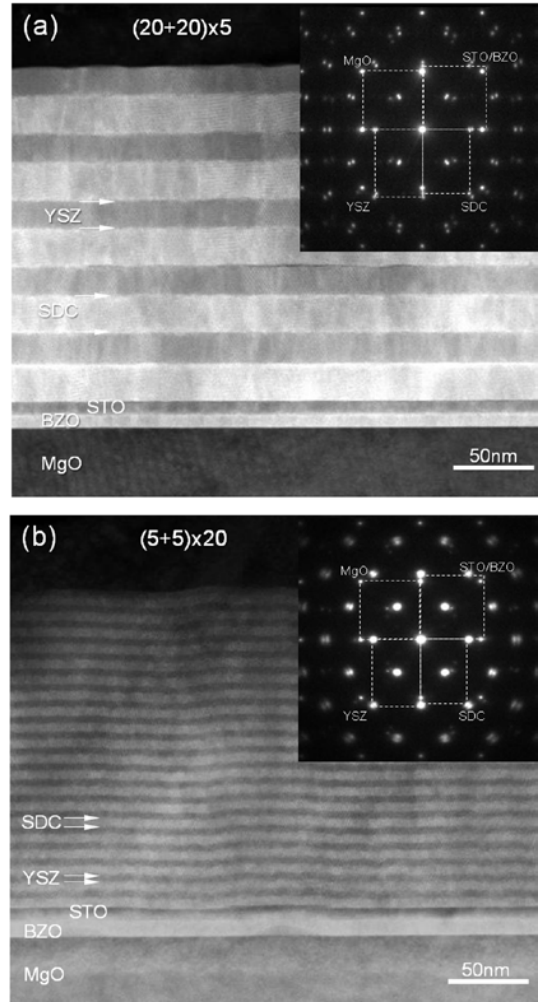

**Figure 2**. Low magnification HAADF-STEM images of a) sample (20+20)×5 and b) (5+5)×20. (SAED) patterns are shown in the insets. YSZ and SDC materials show only spots corresponding to [110] orientation to the direction of the electron beam.

The microstructure of the samples was locally investigated by high-resolution scanning transmission electron microscopy (HR-STEM). Two representative micrographs are reported in Figures 2a and 2b for the samples (20+20)×5 and (5+5)×20, respectively. The selected area electron diffraction (SAED) patterns from both samples show sharps spots from the YSZ and SDC layers corresponding to the [110] zone axes, revealing the epitaxial [001] oriented growth of the layers parallel to [001] orientation of MgO in a good agreement with the XRD data. Highly ordered microstructures were observed with no significant evidence of high angle grain boundaries.

The general characteristics of these samples are very similar to those reported in a previous study.[19] More details on the growth of the BZO-STO double buffer layer can be found in reference [24]. Figures 3a and 3b show HR-STEM images of the interface between YSZ and SDC layers. Even at very high magnification the interfaces of the multilayers appear sharp. The regions in the dashed rectangles in Figures 3a and 3b have been filtered using the [111] spots of the Fast Fourier transforms (FFT) and are shown in Figures 3c and 3d respectively. A high density of misfit dislocations is observed along the YSZ/SDC interface, with Burgers vectors parallel to the interface plane.

From the images, the dislocations were found to occur every 5.5 to 9.5 nm, yielding density of 3.3 to $1.1 \times 10^{16}$ m$^{-2}$, assuming an identical density along both directions parallel with the interface due to the cubic crystal structures. By averaging over 16 dislocations we have found the average spacing between each dislocation to be 6.8±1.2 nm. This is in excellent agreement with the spacing observed between dislocations at the interface of CeO$_2$ films grown on YSZ substrates by PLD,[26] however in our case the spacing was found to vary to a greater extent. Following the same general considerations reported in reference [26] an effective compressive strain in the SDC layer of $0.3^{+0.4}_{-0.6}$% can be estimated, with a tensile strain of the same magnitude expected to occur in the YSZ layers. Hence, the large lattice mismatch of more than 5% between two materials is largely accommodated by the introduction of misfit dislocations, however a degree of residual strain remains.

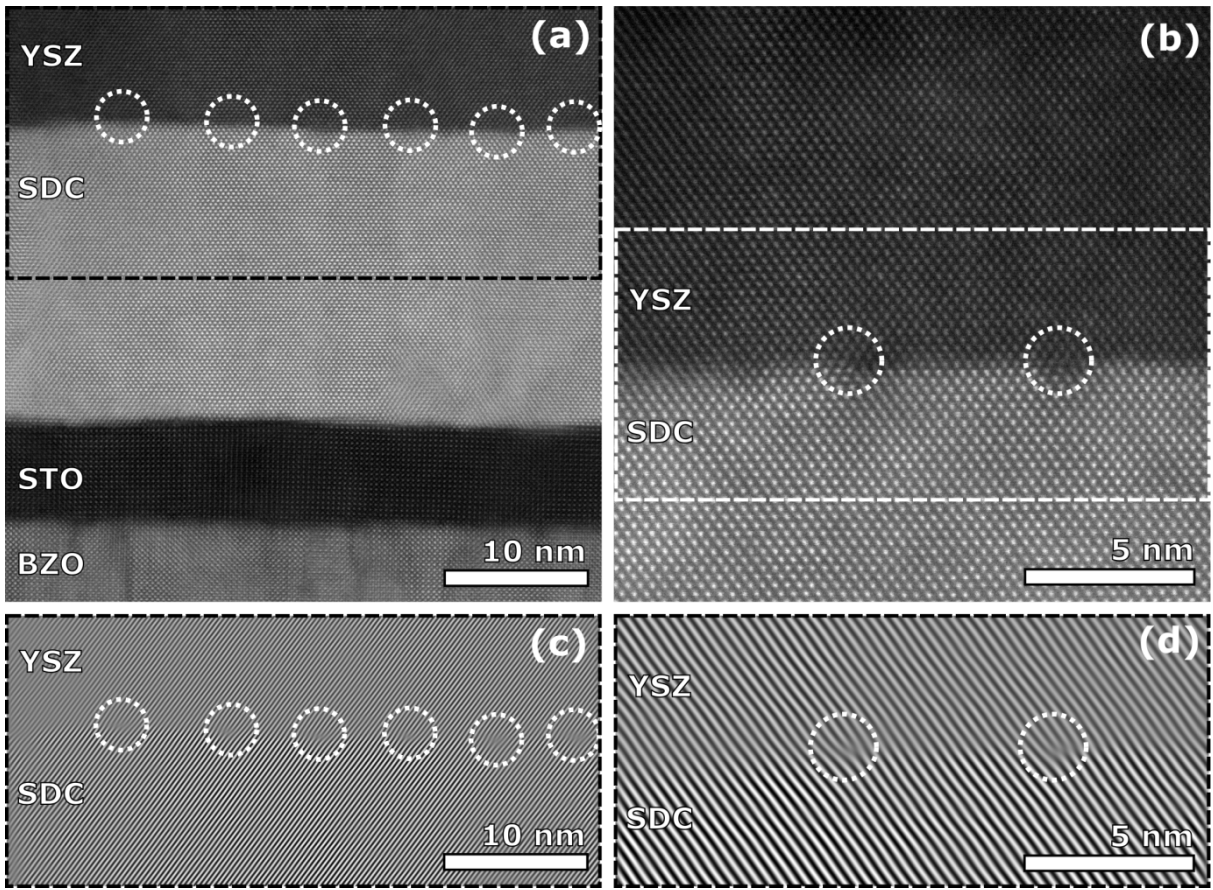

**Figure 3**. HR-STEM images (a, b) of a YSZ/SDC interface in the (20+20)×5 sample. In (c) and (d) FFT filtered images of the regions with the dashed rectangles in (a) and (b) respectively are shown. The misfit dislocations have been marked in both sets of images with dashed circles.

The variation in spacing between dislocations also implies that the strain is not uniform but varies at different regions along the interface, and that the value stated above represents an average. However, it must be noted that due to the limited regions that could be imaged in the TEM sample, the numbers above are not statistically significant and therefore large errors on the strain value have been stated.

In order to assess the chemical intermixing between the SDC and YSZ layers, EDX maps were obtained of a SDC/YSZ interface in the (20+20)×5 sample from a region close to the substrate. As shown in the line profile from an EDX map in Figure 4, no accumulation of either Sm or Y to the interface region was observed. The width of the interface, as marked on the graph is approximately 3 nm. However, the sample was observed to drift by around 0.5 nm over the course of the acquisition, which would act to broaden the width of the interface in the profile. In addition, atomic scale roughness at the interface is also expected to cause broadening of the interface in the profile and hence the value above represents the upper limit of the chemical interdiffusion between the YSZ and SDC layers. This result is expected to be representative for all SDC/YSZ interfaces in the multilayers.

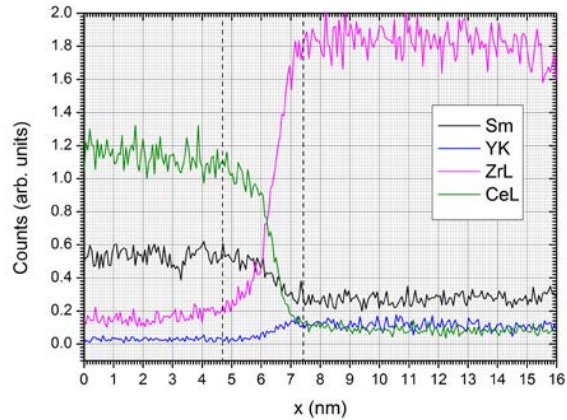

**Figure 4.** Line profile taken from an EDX map of a SDC/YSZ interface in the (20+20)×5 sample. The dashed lines represent the estimated width of the interface.

The SDC/YSZ interfaces were finally analyzed by electron energy loss spectroscopy (EELS) to investigate potential compositional and/or chemical changes. Acceptor dopants have been reported to segregate to surfaces and grain boundaries of doped ceria.[27-28] Also theoretical simulations predict the accumulation of trivalent cations and oxygen vacancies at the dislocations as a consequence of the local expansion of the lattice.[29] Although no substantial accumulation of dopants to the interfacial regions was seen from the EDX analyses, the trivalent cations may include not only Sm but also Ce. A mixture of $Ce^{4+}$ and $Ce^{3+}$ was indeed found along both doped and undoped ceria in $CeO_2$/SDC thin film heterostructures, with the presence of $Ce^{3+}$ was not confined at the interface regions.[30]

The samples (5+5)x20 and (20+20)x5 were analyzed by EELS at the Ce $M_4$ and $M_5$ edges to probe the presence of $Ce^{3+}$ in the multilayers. Experimental spectra were compared with the reference spectra of $CeO_2$ and $CePO_4$ analyzing the peak positions on an energy scale calibrated using Zero Loss Peak (ZLP) and the $M_4$/$M_5$ peak ratio. Fig. 5 shows the maps of the concentration of $Ce^{3+}$ and $Ce^{4+}$ along the SDC layers in the (20+20)×5 (Fig. 5a) and (5+5)×20 (Fig. 5b) samples, respectively. The Ce ions are found predominantly in the 3+ oxidation state within a layer extending about 2 nm from the interface with YSZ. The concentration of $Ce^{3+}$ rapidly decreases beyond the interfacial layer towards the bulk of the SDC layers. At a distance of about 3 - 4 nm from the interface the Ce ions are mainly in the 4+ oxidation state, as expected for SDC. Due to the interfacial roughness the thickness of the $Ce^{3+}$ rich layer is not uniform, thus making a more detailed analysis of its compositional gradient not meaningful.

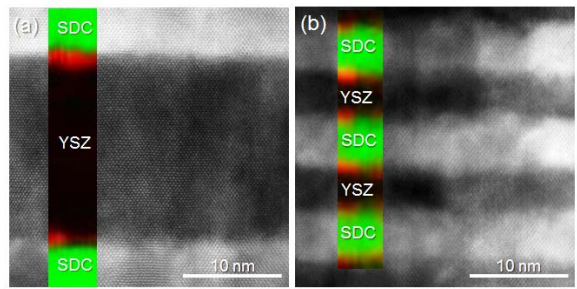

**Figure 5**. EELS measurements of the concentration profile of $Ce^{3+}$ (red) and $Ce^{4+}$ (green) along the ceria layers in the sample a) (20+20)×5 and b) (5+5)×20.

It has been reported in the literature that a very similar result was observed in epitaxial thin films of $CeO_2$ grown on YSZ substrates.[26] For these samples EELS revealed that almost 100% of the Ce ions were reduced from $Ce^{4+}$ to $Ce^{3+}$ near the interface with YSZ and in the first 2 nm. It was also shown that the $Ce^{3+}$ concentration profile decreased rapidly within about 4 nm from the interface.[26] Our finding shows that this effect is independent from the presence of the dopant. Though for the multilayers in this study, the interfacial roughness is much more pronounced than in [26] where a YSZ single crystal substrate was used, we may assume as a first approximation that in our systems the chemical profile is locally the same as that described in detail in [26].

We note that the thickness of the reduced layer is similar to the accuracy of the EDX measurement of the chemical line profile of Fig. 4 implying that we cannot rule out a significant intermixing on the two materials within a 2 nm layer at their interface. It is known that the solution of Zr in $CeO_2$ can significantly decrease the reduction energy of the $Ce^{4+}$ ions.[31] However, this effect becomes significant only at very low oxygen partial pressures ($< 10^{-10}$ atm), thus we conclude that the formation of a ceria-zirconia solid solution does not support the high density of $Ce^{3+}$ observed.

Previously it has been reported that $CeO_2$ can be reduced *in situ* during TEM measurements due to interactions with the electron beam.[32] However in this work this is unlikely to be the case due to the following reasons: Firstly, no change in the samples were observed over the course of the STEM measurements and no evidence of oxygen vacancy ordering in the images or diffraction patterns were observed such as that seen during $CeO_2$ reduction under the electron beam in the literature. Finally, the $Ce^{3+}$ observed by EELS in this study is found only locally at the interfaces between the YSZ and SDC and not in the center of the layers despite the uniform scanning of the electron beam over the entire region. This is in contrast to those experiments where Ce reduction is induced *in situ* during the TEM measurement, where the Ce reduction is more uniform in the regions imaged.

It has also been recently reported that cerium reduction in $CeO_2$ is facilitated by high strain, either compressive (-5.6%) or tensile (2.1%).[33] Nevertheless the in-plane strain required to favor the cerium reduction is much higher than that measured in our samples (0.3%), and this effect therefore should not give significant effect. So far, the driving force for the reduction of Ce ions in $CeO_2$ (or SDC) at the interface with YSZ remains unclear and certainly deserves further investigation.

**Electrical characterization**

Ceria reduction can be described by the following reaction (in Kröger - Vink notation):

$$O_O^x + 2Ce_{Ce}^x \rightleftarrows V_O^{\cdot\cdot} + 2Ce_{Ce}' + \frac{1}{2}O_2 \quad . \tag{1}$$

It is generally accepted that electrons are strongly localized at the cerium atoms,[34-35] and are charge compensated by oxygen vacancies.[36-37] Therefore for electroneutrality reasons:

$$\left[Ce_{Ce}'\right] = 2\left[V_O^{\cdot\cdot}\right] . \tag{2}$$

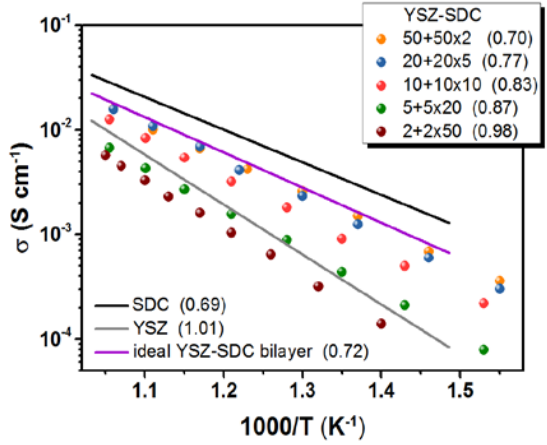

**Figure 6.** Conductivity measurements of the five heterostructures (circles). The dark and light gray solid lines show the linear fit of the conductivities of fully relaxed epitaxial films of SDC and YSZ, respectively. The purple solid line is the theoretically expected conductivity of an SDC+YSZ bilayer calculated as $(\sigma_{SDC}+\sigma_{YSZ})/2$. In the legend the activation energies in eV are reported in brackets.

The oxygen vacancy concentration is set by the majority of negatively charged defects, which in this case are reduced cerium atoms. Following reference [26], the presence of $Ce^{3+}$ near the interface can be compensated either by oxygen vacancies within the ceria layer, at the interface or in the YSZ layer. All these mechanisms have potential implications on the conductive properties of the interfacial region.

The electrical characterization was performed by impedance spectroscopy using Pt electrodes. The conductivity ($\sigma$) measurements are fitted to the linearized Arrhenius equation $\ln(\sigma T) = \ln(\sigma_0) - \frac{E_a}{k_B T}$, where $\sigma_0$ is the pre-exponential factor (proportional to the density of charge carriers, the attempt frequency for hopping, and the distance between hopping sites), $E_a$ the activation energy for oxygen-ion migration, $k_B$ and T the Boltzmann constant and the absolute temperature, respectively. Figure 6 shows the Arrhenius plot of the conductivity of the SDC-YSZ heterostructures fabricated for this work, as well as the linear fit of the conductivity of 200 nm thick epitaxial films of SDC and YSZ used as references for this study (dark- and light-gray solid lines, respectively). These two films show the typical bulk conductivity of SDC [38-39] and YSZ [40-41] respectively. Figure 6 also reports the simulated conductivity of an ideal fully relaxed SDC-YSZ bilayer (purple solid line) calculated as $(\sigma_{SDC} + \sigma_{YSZ})/2$ using the bulk conductivities of the two materials assuming the two layers as two resistances in parallel configuration.

As a first observation, the conductivity and activation energy of the sample (50+50)×2 showing the fully relaxed structure for both SDC and YSZ is very similar to the conductivity calculated for the ideal bilayer. This result is an important internal validation and shows that, as expected, in the case of thick layers any potential interface effect is not measurable and that the two materials contribute independently to the total conduction mechanism each with its own bulk conducting properties.

Reducing the thickness of the layers and as a consequence enhancing the contribution of the interfacial regions to the total conductivity, the conductivity of the heterostructures gradually decreased showing progressively larger activation energies and larger pre-exponential factors that approach values very similar to those of YSZ. Indeed the total conductivity of the sample (2+2)×50, which is calculated taking into account the total thickness of the heterostructure, is approximately a factor of two smaller than the YSZ bulk conductivity and shows a similar activation energy. Considering that in this sample the YSZ layers account for half of the total thickness of the conductor we conclude that by reducing the layer thickness the oxygen ion transport in these heterostructures is progressively confined along the YSZ layers with conducting properties similar to bulk. For the smallest thickness of 2 nm the SDC films act *de facto* as insulating layers.

We can conceive of three possible ways the conductivity could be modulated in our systems as the thickness of the layers is decreased: microstructural/morphological changes, strain effects and compositional or chemical changes.

It has been reported that grain sizes in $CeO_2$ films grown by PLD can be smaller in regions closer to the substrate,[39] and it is well known that the specific grain boundary resistance and activation energy is much larger than the grain interior for acceptor-doped ceria. Therefore, an increase in the density of grain boundaries with decreasing layer thickness could yield changes in the conductive properties as observed for these samples. However, from the TEM analysis of these samples, no evidence for a significant density of high-angle grain boundaries or for changing grain size with layer thickness was observed. Additionally, in a review of doped ceria thin films,[42] it has been noted that decreasing the grain size for nano-granular thin films typically results in a decrease in the activation energy of the total film conductivity. Hence, we suggest that microstructural/morphological changes are unlikely to be the cause of the changes in the conductive properties of the multilayers.

The second possibility is due to strain effects in the multilayers. Kushima and Yildiz[17] theoretically predicted a linear increase of the diffusion coefficient of YSZ from compressive to tensile strain up to a maximum tensile strain of around 3%. Using multilayered heterostructures fabricated by coupling YSZ to an insulating phase, Korte[43] and Aydin[21] (and co-workers) experimentally investigated the effect of strain along the YSZ layers. As a consequence of a lower activation energy, a conductivity enhancement by a factor of about 3.5 induced by a 0.8% tensile strain was reported at 500°C. A similar effect of tensile strain on the conductivity was also recently observed for doped ceria;[14] where the activation energy for oxygen ion transport decreased by 0.05 eV as a consequence of a tensile strain of about 0.35% leading to twofold larger conductivity at 350°C.

SDC/YSZ heterostructures grown on STO-buffered MgO substrates were studied in reference[23]. In that case, a large conductivity enhancement with progressively larger activation energy was observed by increasing the number of bilayers at constant total thickness, as well as by decreasing the single layer thickness keeping constant the number of bilayers. In spite of a sample design very similar to that used in the present study, the XRD analysis revealed a different microstructural environment. Large (in-plane tensile) lattice distortions were found for the YSZ layers, while the SDC layers maintained the almost relaxed lattice parameter. The enhanced conductivity was ascribed to the tensile strain along the YSZ layers while the contribution of the SDC layers was assumed to remain constant. It is not clear why the activation energy increased with increasing tensile strain, since the opposite trend is expected.

Azad and coworkers reported the electrical characterization of (111)-oriented, 12 at.% GDC-GSZ multilayers grown on sapphire.[22] Lower activation energies were measured by decreasing the thickness of each layer down to a minimum thickness of about 15 nm (at constant total thickness of 155 nm). Further decreasing the layer thickness, the activation energy increased. The reported conductivity values were in all cases larger than the typical conductivity of YSZ bulk, but they appear smaller than the expected bulk conductivity of Gd-doped ceria. We may assume that the effect of tensile strain (lower activation energy) along the GSZ layers dominates the total conductivity down to a minimum thickness of about 15 nm, while the effect of compressive strain (larger activation energy) along the ceria layers became more significant for smaller thicknesses.

In the present study, the progressively lower conductivity and larger activation energy (with values approaching those of bulk YSZ) measured while decreasing the bilayer thickness of SDC/YSZ heterostructures would suggest that the effect of tensile strain on YSZ is negligible while the compressive strain of SDC dramatically affects the conductivity. In our previous study,[19] the conducting properties of $CeO_2$-YSZ heterostructures fabricated in a similar way were investigated. Down to a minimum thickness of 5 nm, the YSZ layers did not show any significant effect due to the residual tensile strain, which was most probably of a similar extent as that estimated here. The heterostructures fabricated for this work show very similar structural and morphological characteristics as those reported in [19], thus one may assume that the YSZ layers preserve conducting properties similar to bulk also in these multilayers. However, assuming only the strain as the origin of the observed effect we should conclude that the compressive or tensile strain has a very different effect on the conductivity of doped ceria. On one hand a tensile strain of 0.35% induced a relatively small reduction of the activation energy resulting in a twofold increased conductivity at 350°C.[14] On the other hand, a compressive strain of a similar extent almost completely suppresses the charge carrier migration along the layer. We thus conclude that strain effects cannot satisfactorily describe our experimental results.

The final possibility we suggest to rationalize our experimental results is due to compositional or chemical changes. The almost complete reduction of the Ce ions at the interface, as in [26], would imply the presence of an interfacial layer with a stoichiometry close to that of cerium sesquioxide, which is stable as a-$Ce_2O_3$ with a hexagonal crystal structure. Since XRD, RHEED and HR-TEM analyses showed no indication of such a phase change the only possibility is the stabilization of the cubic c-$Ce_2O_3$ (bixbyite). Stetsovych and co-workers[44] experimentally demonstrated the possibility of stabilizing highly ordered nanometric thin films of the cubic bixbyite c-$Ce_2O_3$. The lattice parameters of the Ce and Sm cubic bixbyite are about 11.16 and 10.95 Å [45-46] yielding a lattice misfit of about 2.7 and 0.8% with $CeO_2$ respectively. The stabilization of interfacial layers of cubic Ce and Sm bixbyite with composition close to $Ce_{0.85}Sm_{0.15}O_{1.5}$ is consistent with the XRD, RHEED, HR-TEM, EDX and EELS analyses. It is worth noticing that all these measurements suggest that in these heterostructures such a high deviation from stoichiometry is stabilized in air, while typically it can be achieved only at very low oxygen partial pressures ($pO_2$) and high temperatures. For instance for reduced ceria $CeO_{2-x}$, values of $x \approx 0.2$ are obtained at $pO_2 \approx 10^{-23}$ atm at 800°C;[36] while in Sm- or Gd-doped ceria $x \approx 0.2$ has been obtained at $pO_2 \approx 10^{-16}$ atm at 1000°C[47] and $x \approx 0.1$ has been reported at $pO_2 \approx 10^{-25}$ atm at 700°C.[48]

This leads to the question: what are the expected conductive properties of such a strongly reduced ceria layer?

Compared to $CeO_2$ this highly ordered thin film of a normally unstable polymorph of $Ce_2O_3$ contains 25% of perfectly ordered oxygen vacancy clusters consisting of four oxygen vacancies.[45] Along these layers the clustering and ordering of the oxygen vacancy sublattice would result in a dramatic drop of the oxygen-ion conductivity leading to a progressive confinement of the oxygen ion transport along the YSZ layers with decreasing layer thickness in agreement to the experimental results.

Oxygen ions are not the only type of charge carriers that can potentially contribute to the total conductivity. In general, the reduction of $Ce^{4+}$ to $Ce^{3+}$ enhances the electronic conductivity which, due to the high mobility of electrons, significantly increases the total conductivity.[49] It may be argued that the topmost YSZ layer (which is a good electronic insulator) makes the IS measurements insensitive to the electronic conductivity. However, the aspect ratio of the electrode area (about 20 mm$^2$) over film thickness (200 nm) has to be considered: The aspect ratio of across-plane and in-plane differ by order of magnitudes (about 1x10$^{-8}$). Also taking into account the very low electronic conductivity of YSZ, the large thickness/area ratio of the electrodes would make the conductance sufficiently high to detect the in-plane electronic current of SDC. Therefore, it is highly unlikely that the measured electrical conductivity does not also include the electronic contribution. To confirm that any possible electronic contribution to the conductivity can be detected, the electrical conductivity measurement was repeated for one of the (10+10)×10 samples at 700°C in an oxygen partial pressure (pO$_2$) estimated to be between 10$^{-23}$ and 10$^{-25}$ atm. In these environmental conditions the ceria layers are reduced and the conductivity is predominantly electronic in the range of 0.8 – 1 S cm$^{-1}$ (to be compared to the ionic conductivity 0.07 S cm$^{-1}$ measured in air in the same temperature range).[50] On the contrary, the YSZ conductivity remains unperturbed and shows the same values of about 0.01 S cm$^{-1}$ measured in air.[51] With our sample we measured a total conductivity of ≈ 0.018 S cm$^{-1}$ at 700°C in air. No significant changes were detected in N$_2$ while adding 5% H$_2$ and reducing the pO$_2$ down to the range between 10$^{-23}$ and 10$^{-25}$ atm the conductivity increased to ≈ 0.3 S cm$^{-1}$.

This measurement clearly shows that with our sample design the measured conductivity includes also the electronic contribution. Therefore we conclude that the electronic contribution to the conductivity is negligible during the conductivity measurement in air. We stress further that the results of the IS measurement in air and in reducing atmosphere represent two very different situations: in the first one cerium is reduced over a thickness of only a few nanometers close to the interface, in the second one the GDC layers are reduced over the entire thickness (the degree of cerium reduction in the two cases is different).

As discussed above, the experimental results summarized in Fig. 6 strongly suggest that the electronic conductivity in our sample (if present) is negligible compare to the ionic contribution. The next question is: is this consistent with the hypothesized stabilization of the cubic bixbyite c-$Ce_2O_3$ structure at the interfaces? Indeed, a very high concentration of reduced Ce ions is expected to suppress the electronic conductivity. In fact, the electronic conductivity at a certain temperature in reduced ceria is described as small-polaron hopping ($Ce'_{Ce}$) according to $\sigma_{el} = |v|e\,[Ce'_{Ce}] \cdot \mu(1-f)$,[35] where $v$ is the valence of the charge carrier (+1 for $Ce'_{Ce}$), $e$ the electronic charge, $[Ce'_{Ce}]$ the concentration, and $\mu(1-f)$ is the reduced mobility which takes into account the lattice site fraction already occupied by $Ce^{3+}$ since the polaron needs a free $Ce^{4+}$ site ($Ce^{\times}_{Ce}$) to migrate to. In comparison with the literature,[26] for our heterostructures we may assume that the majority of the cerium crystallographic sites at the interface with YSZ are occupied by ($Ce'_{Ce}$) or ($Sm'_{Ce}$). In such a case, $f \approx 1$ and the electronic contribution to the total conductivity would be small.

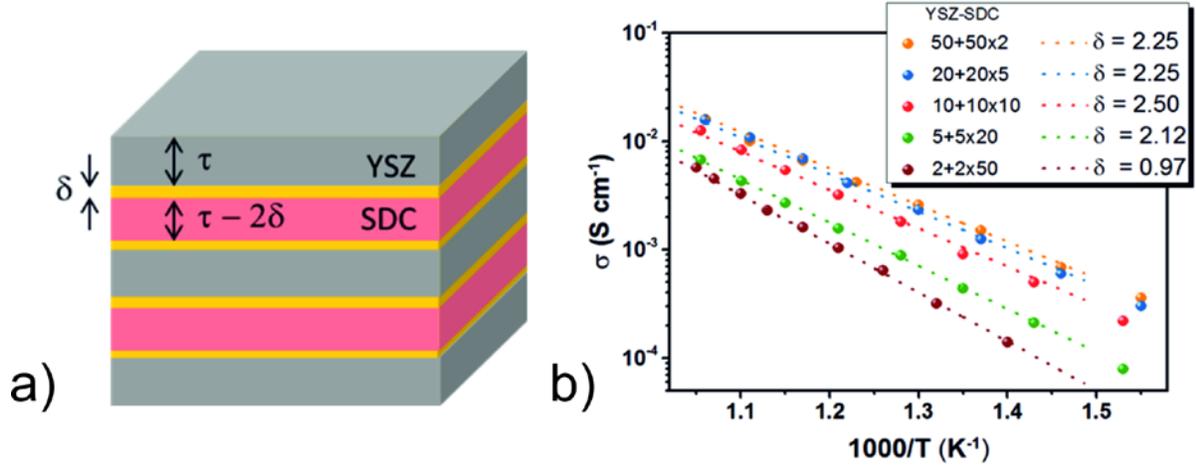

**Figure 7**. (a) Scheme of the model system used to fit the experimental data. (b) Best fit of the experimental data (dot lines). The best values of δ (in nm) are reported in the legend.

To summarize, the stabilization of the cubic Ce/Sm bixbyite structure in a layer about 2 nm thick at each interface with YSZ is expected to suppress both the ionic and electronic conductivity along these layers which would thus behave as insulating interlayers. In view of these considerations, as a first approximation we fit the measured conductivity to the expected conductivity of a simplified model consisting of an ideal SDC/YSZ multilayer with an insulating layer of thickness δ along the SDC films at all interfaces with YSZ, as shown in Fig. 7a. The conductivity of such an ideal sample is given by

$$\sigma = \frac{1}{2}\left[\sigma_{YSZ} + \sigma_{SDC}\left(1 - \frac{2n+1}{n}\frac{\delta}{\tau}\right)\right]$$

where $\sigma_{YSZ}$ and $\sigma_{SDC}$ are the bulk conductivities of the two materials (Fig. 6), $n$ is the number of bilayers (2, 5, 10, 20, and 50), and $\tau$ is the thickness of the YSZ layers (50, 20, 10, 5, and 2 nm).

The dotted lines in Fig. 7b show the fit to the experimental data calculated using δ as a free parameter. The best fit of the experimental data was obtained by the minimization of the $\chi^2$ value (normalized to the number of degree of freedom) while varying δ. The legend of Fig. 7b also reports the estimated values of δ in nm.

The described model is certainly oversimplified; however it shows a remarkable agreement with the experimental data for all layer thicknesses suggesting the presence of an almost insulating ceria layer with a thickness of about 2 nm (the same thickness where Ce was found primarily in the $3^+$ oxidation state) at each interface with YSZ.

For the sample (2+2)×50 the best fit of the experimental data is obtained using a value of δ ≈ 1 nm, which is consistent with all other samples considering that in this case the layers' thickness is 2 nm. In other words, according to the model, in the sample (2+2)×50 the entire thickness of the SDC layers does not contribute to the conductivity. This is consistent with the result shown in Fig. 6. It is interesting to note that the slightly smaller activation energy measured for the sample (2+2)×50 compared to bulk YSZ is in good agreement with the expected effect of an in-plane tensile strain of the conductor of the order of 0.3%,[14, 17, 21, 43] which is the value estimated by HR-TEM for these heterostructures.

**Conclusions**

SDC/YSZ epitaxial heterostructures were fabricated to investigate the properties of the interface between the two materials. We found the presence of a $Ce^{3+}$-enriched layer about 2 nm thick at the ceria side of each interface. The electrical characterization showed that these interlayers have very small ionic and electronic conductivity compared to bulk YSZ and that lowering the layer thickness in the heterostructures progressively confines the ionic transport only along the YSZ layers. For the smallest layer thickness of 2 nm, the SDC layers do not contribute to the conductivity, and the oxygen ion transport only occurs through the YSZ.

The combination of electrical, structural, morphological and compositional analyses points to the stabilization of 7.5 mol.% Sm doped cubic bixbyite ceria with a composition close to $Sm_{0.15}Ce_{0.85}O_{1.5}$ along the $Sm_{.15}Ce_{0.85}O_2$ (SDC) layers at the interfaces with YSZ. Along these layers the ordering of the oxygen vacancy sublattice and the almost complete reduction of the Ce ions hinder both the ionic and electronic conductivities.

## AUTHOR INFORMATION

**Corresponding Author**

* daniele.pergolesi@psi.ch

**Author Contributions**

The manuscript was written through contributions of all authors. All authors have given approval to the final version of the manuscript.

**Funding Sources**

This research was supported by the NCCR MARVEL, funded by the Swiss National Science Foundation

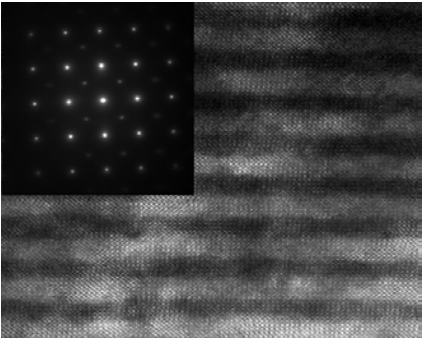

For Table of content only